\documentclass[12pt,a4paper]{article}
\usepackage{graphicx}
\usepackage{times}
\usepackage{pdflscape}
\title{\bf Probing the Galactic s-process nucleosynthesis using barium stars 
}

\author{Shejeelammal Jameela$^1$, Aruna Goswami$^1$\\
\vspace{0.5cm}\\
\normalsize $^1$ Indian Institute of Astrophysics, Koramangala 2nd Block, 
Bangalore - 560034, India} 

\date{\mbox{}}

\begin{document}
\maketitle
\setcounter{page}{1001}
\pagestyle{plain}
    \makeatletter
    \renewcommand*{\pagenumbering}[1]{
       \gdef\thepage{\csname @#1\endcsname\c@page}
    }
    \makeatother
\pagenumbering{arabic}

\def\bull{\vrule height .9ex width .8ex depth -.1ex}
\makeatletter
\def\ps@plain{\let\@mkboth\gobbletwo
\def\@oddhead{}\def\@oddfoot{\hfil\scriptsize\bull\quad
"2nd Belgo-Indian Network for Astronomy \& astrophysics (BINA) workshop'', held in Brussels (Belgium), 9-12 October 2018 \quad\bull}
\def\@evenhead{}\let\@evenfoot\@oddfoot}
\makeatother

\def\beginrefer{\section*{References}
\begin{quotation}\mbox{}\par}
\def\refer#1\par{{\setlength{\parindent}{-\leftmargin}\indent#1\par}}
\def\endrefer{\end{quotation}}

{\noindent\small{\bf Abstract:} 
The origin and evolution of neutron-capture elements in our Galaxy is poorly understood. 
In this work, we have attempted to understand the Galactic s-process nucleosynthesis using barium stars as probes.
We have performed high resolution spectroscopic analysis for three 
 barium stars HD 36650, HD 207585 and HD 219116. 
The analysis of HD 219116 is based on HCT/HESP data at a resolution 
of 60,000 covering a wavelength region 3530-9970 {\rm \AA}. For HD 36650, 
we have used the FEROS spectrum at a resolution of 48,000 covering the 
wavelength region 3520-9200 {\rm \AA}. For HD 207585, the spectrum
from VLT/UVES at a resolution of 48,000 covering the wavelength 
region 3290-6650 {\rm \AA} is used. Our analysis clearly shows that 
the surface chemical composition of these barium stars are enriched 
by s-process nucleosynthesis products coming from their former companion 
low-mass AGB stars. 
A discussion on the distriburtion of abundance ratios  based on  the 
existing nucleosynthesis theories is presented. 
}
\vspace{0.5cm}\\


\section{Introduction}
All the elements beyond Fe-peak are produced through the slow and 
rapid neutron-capture processes. The Asymptotic Giant Branch (AGB) stars 
are the predominant sites for the slow neutron-capture (s-process)
nucleosynthesis. Low- and intermediate- mass stars 
evolve  through the AGB  phase of stellar evolution. 
The s-process occurs at 
low neutron densities of $N_{n} \approx 10^{7} - 10^{10}$ 
neutrons/cm$^{3}$ (Busso et al. 1999) on a timescale much slower than the 
 $\beta$-decay rate of unstable isotopes. The important neutron sources 
for this process  are $^{13}$C($\alpha$, n)$^{16}$O, for the stars 
with initial mass $\leq 3M_{\odot}$, and $^{22}$Ne($\alpha$, n)$^{25}$Mg, 
for the stars with initial mass $\geq 4M_{\odot}$.
The dominant isotopes of Sr, Zr, Nb, Ba, La and Pb are produced 
through s-process.  The isotopes such as $^{86}$Sr, $^{96}$Mo, $^{104}$Pd, 
$^{116}$Sn  are produced only by the  s-process. Besides these 
heavy elements, AGB stars are important contributors of carbon and 
 nitrogen in the Galaxy. About a third of the total carbon in the Galaxy 
is found to be produced  in AGB stars (Dray et al. 2003). Major producers 
of $^{14}$N in the Galaxy are the intermediate-mass AGB stars with 
Hot-Bottom Burning (Henry et al. 2000). The products of the nucleosynthesis 
from the  interiors are brought to the surface of the AGB star through the 
Third Dredge Ups and these materials are  expelled to the ISM through 
stellar winds. The understanding of the AGB phase and nucleosynthesis 
 is important to study the role of low- and intermediate-mass stars 
in the chemical enrichment of the Galaxy. 

However, observations show that the neutron-capture elements in the 
Galaxy show a large scatter at a given metallicity. The observed scatter 
in the abundances of these elements with respect to metallicity suggest 
that the ISM from which the stars are formed may not be mixed 
homogeneously or same element may have different origins. 
The analysis of Cayrel et al. (2004) shows that the elements with 
Z $<$ 30 show very little scatter, discarding the  claim for 
an inhomogeneous ISM. Detailed chemical analyses of stars 
with the atmosphere enriched by the neutron-capture elements can shed 
light on the origin and chemical evolution history of these elements 
in the Galaxy. Metal-deficient Barium stars constitute useful 
candidates for conducting such studies. 

Barium (Ba II) stars belong to a family of peculiar G and K spectral 
types which were first identified by Bidelman \& Keenan (1951). 
They are mostly in the giant phase of stellar evolution.
Their surface chemical composition  exhibits overabundance of elements 
heavier than iron. An important characteristic property of these stars 
is that  C/O $<$ 1 (Barbuy et al. 1992, Allen \& Barbuy 2006, 
Drake \& Pereira 2008, Pereira \& Drake 2009). They exhibit abnormally 
strong lines of Ba II at $\lambda$ 4554 {\rm \AA}, 
Sr II at $\lambda$ 4077 {\rm \AA}, as well as enhanced CH, CN and 
C$_{2}$ molecular bands. They are low radial-velocity objects and 
generally members of the Galactic disk.  A fraction of them show mild 
metal deficiency. The analysis of Luck \& Bond (1991) shows that C$_{2}$ 
molecular bands in metal-deficient Ba stars are not strong as 
compared to the classical Ba stars.
 
All giant barium stars are found to be in binaries with a now 
invisible white dwarf companion  (McClure et al. 1980, McClure 1983, 
1984, McClure \& Woodsworth 1990, Udry et al. 1998a,b, Jorissen et al. 2019). 
A generally accepted scenario that explains 
the observed high abundances of neutron-capture elements is a binary 
picture. These stars are believed to have received via binary 
mass-transfer mechanisms the nucleosynthesis products of the 
companion star produced during its AGB phase of evolution. Hence, 
the chemical composition of this class of objects can be used to trace the
AGB nucleosynthesis at their corresponding metallicity. 

In section 2, we presented the spectroscopic methodology. Results and 
discussions are presented in section 3. Conclusions are drawn in 
section 4.

\section{Spectroscopic Methodology}
\subsection{Object selection}
With the aim to understand the Galactic s-process nucleosynthesis using 
barium stars as probes,  we have selected a sample  of  
Ba star candidates from various sources (L\"u 1991, 
Bartkevicius 1996). The results from our analysis of the three stars 
HD 36650, HD 207585 and HD 219116 are presented  in this work. 

\subsection{Data acquisition.}
The high resolution spectrum for HD 219116 is acquired using 
HESP attached to the Himalayan Chandra Telescope (HCT) at a resolution
of  R$\sim$60,000. The wavelength coverage of this spectrum spans from
3530 to 9970 {\rm \AA}. A VLT/UVES spectrum at a resolution of R$\sim$48,000
is used for HD~207585. The wavelength coverage of this spectrum is 
3290-6650 {\rm \AA}. The analysis of  HD 36650 is based on a FEROS spectrum
at a resolution  of R$\sim$48,000, and covers the  
wavelength region 3520-9200 {\rm \AA}. All the spectra have S/N $>$ 30. 
The basic data of the program stars are given in  Table \ref{basic data}.

\begin{table}
\begin{center}
\centering
\caption{\textbf{Basic data for the program stars.}\label{basic data}}
\begin{tabular}{|l|l|l|l|l|l|} 
\hline 
             &              &                &               &        &                     \\
   Star name & RA(2000)     & DEC(2000)      & Spectral type & V      & Parallax(mas)    \\
             &              &                &               &        &                                \\
\hline  
   HD 36650 (FEROS)  & 05 27 42.92  & -68 04 27.16   & K0 III        & 8.79   & 2.655$\pm$0.027   \\
   HD 207585 (UVES)  & 21 50 34.71  & -24 11 11.68   & G2 II         & 9.78   & 5.315$\pm$0.407   \\
   HD 219116 (HESP)  & 23 13 30.24  & -17 22 08.71   & G8 III        & 9.25   & 1.584$\pm$0.044   \\
  \hline
\end{tabular}
\end{center}
\end{table}

\subsection{Data Reduction}
The data reduction is done using the basic tasks in Image Reduction and Analysis Facility (IRAF) software.
 An image frame is first trimmed to get an area that contains only useful data. 
 The image frame is then bias subtracted followed by the flat correction. 
 Different orders from the spectrum are then extracted using the task 'APALL'.
 The spectrum is then wavelength-calibrated using a Th-Ar arc spectrum. The wavelength calibrated image 
 is then corrected for dispersion and continuum-fitted for further analysis.

\subsection*{Data Analysis}
 Absorption lines due to different elements are identified by comparing closely 
 the spectra of the program stars with the Doppler-corrected spectrum of the star Arcturus. 
 The log $gf$ and  excitation potential values are taken from the Kurucz database of 
 atomic line lists. The equivalent width of good clean  spectral lines 
are measured using 
 various tasks in IRAF. A master line list including all the elements was generated. 

An initial model atmosphere was selected from the Kurucz grid of model atmospheres \\
(http://cfaku5.cfa.hardvard.edu/) using the photometric temperature
estimate and the log g value for giants/dwarfs. A final model atmosphere was adopted 
by an iterative method. The effective temperature is determined by the method of excitation equilibrium, 
micro-turbulent velocity by equivalent width balance and surface gravity by ionization balance of Fe I and Fe II lines. 

The log \textit{g} value is also calculated using the parallax taken from GAIA DR2 archive.
The bolometric correction, BC, is determined using the empirical calibration that links BC with the 
effective temperature of the star from Alonso et al. (1999). 
The interstellar extinction is determined as described by Chen et al. (1998). 
The derived atmospheric parameters of the program stars are given in  Table \ref{atmospheric parameters}.
\begin{table}
\centering
\caption{\textbf{Derived atmospheric parameters of the program stars}} \label{atmospheric parameters}
\begin{tabular}{|l |l |l |l |l |l |}
\hline\hline
Star name  & $T_{eff}$($\pm$100)    & log \textit{g}($\pm$0.2)       & log \textit{g}       &  [Fe/H] & $\zeta$($\pm$0.1)   \\ 
           &         (K)            &     (spectroscopic)            & (parallax)           &         &    ($kms^{-1}$)    \\[0.2ex]
\hline
HD 36650  & 4880  & 2.40  & 2.78$\pm$0.01  & $-$0.02$\pm$0.13  & 1.30  \\
HD 207585 & 5800  & 3.80  & 3.90$\pm$0.05  & $-$0.38$\pm$0.12  & 1.00  \\
HD 219116 & 5050  & 2.50  & 2.68$\pm$0.02  & $-$0.45$\pm$0.11  & 1.59   \\
\hline
\end{tabular} 
 \end{table}
 
For  most of the elements, abundances are  determined from the measured 
equivalent width of 
lines of the neutral and ionized atoms using the adopted model atmosphere. For the analysis, 
the radiative transfer code MOOG (Sneden 1973) is used with the assumption of
Local Thermodynamic Equilibrium (LTE).
For the elements- Sc, V, Mn, Co, Cu, Ba, La and Eu, showing hyper-fine splitting and for
the elements with only single clean line in the spectra and for the molecular bands, 
spectrum synthesis from MOOG has been used to find the abundances. 
The hyper-fine components of Sc, V and Mn are taken from Prochaska \& 
McWilliam (2000), 
Co and Cu from Prochaska et al. (2000), Ba from McWilliam (1998), La 
from Jonsell et al. (2006) 
and Eu from Worely et al. (2013). The solar abundance values of 
the elements are taken from Asplund et al. (2009).

\begin{figure}
\centering
\includegraphics[scale=0.4]{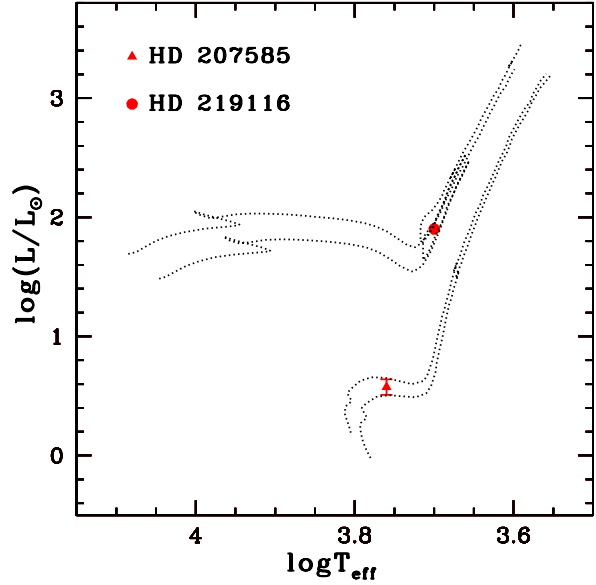}
\caption{\small{Location of HD 219116 and HD 207585 in the 
Hertzsprung-Russell diagram. The evolutionary tracks are 
shown for 1.0, 1.1, 2.2 and 2.5$M_{\odot}$ 
from bottom to top for z = 0.008.}}\label{HR1}
\end{figure}

\begin{figure}
\centering
\includegraphics[scale=0.4]{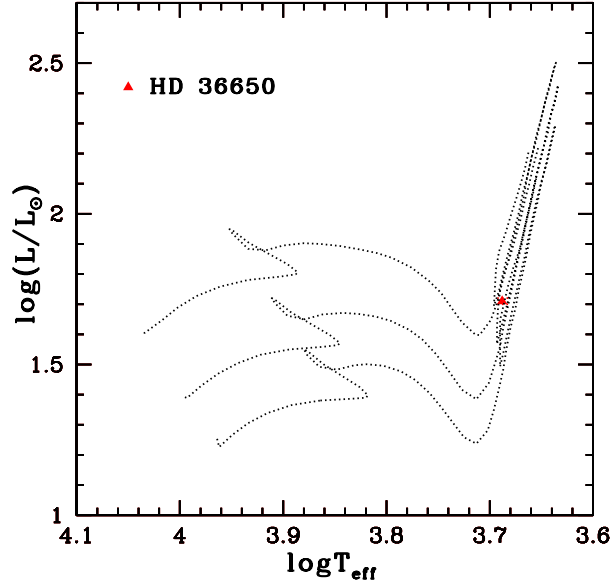}
\caption{\small{Location of HD 36650 in the Hertzsprung-Russell 
diagram. The evolutionary tracks are shown for 2.0, 
2.2 and 2.5$M_{\odot}$ from bottom to top for z = 0.019.}}\label{HR2}
\end{figure}

The stellar evolutionary stage of the program stars is found 
from their location in the Hertzsprung-Russell diagram 
(Girardi et al. 2000 data base of evolutionary tracks) using 
the luminosity and spectroscopic 
temperature estimates. The position of the program stars in the HR diagram 
is shown in Figures \ref{HR1} and \ref{HR2}.

\section{Results  and Discussion}

\begin{figure}
    \centering
\includegraphics[width=0.73\linewidth,height=0.73\textwidth]{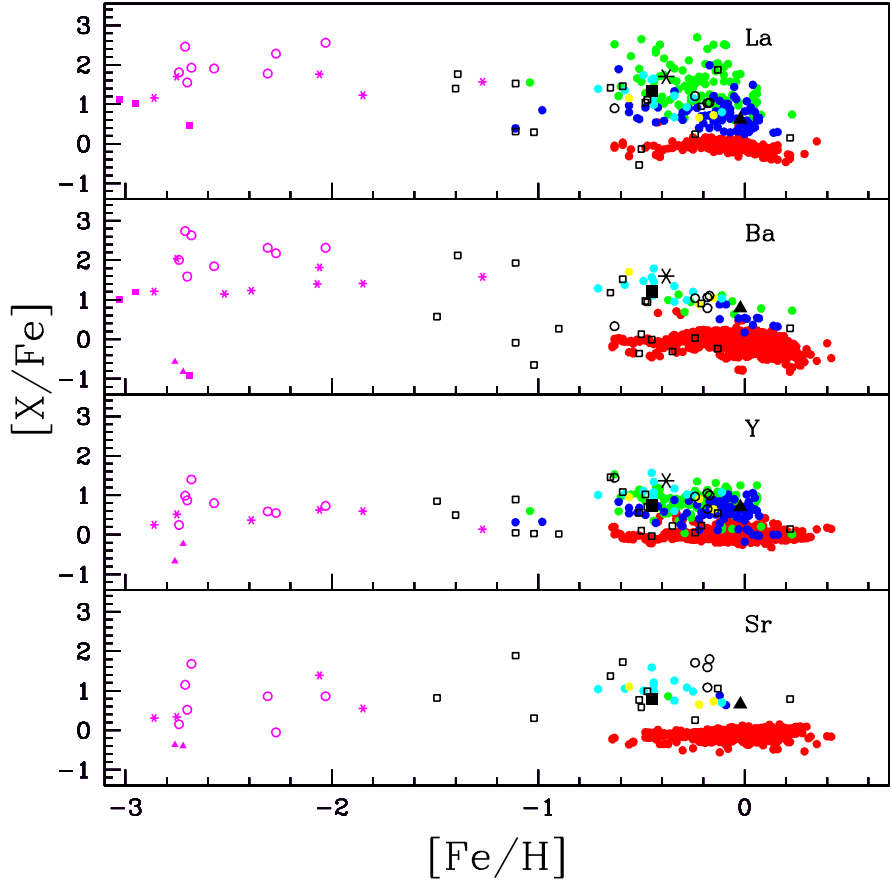}
\caption{\small{Abundance ratios of heavy elements Sr, Y, Ba, La observed 
in the program stars as a function of metallicity [Fe/H]. Red circles 
represent normal giants from literature (Luck \& Heiter 2007). Green, blue, 
cyan, yellow circles represent strong Ba giants, weak Ba giants, 
Ba dwarfs, Ba subgiants respectively from literature (de Castro et al. 
2016, Yang et al. 2016, Allen \& Barbuy 2006). Magenta stars, filled 
squares, open circles, filled triangles represent the CEMP-s, CEMP-r, 
CEMP-r/s, CEMP-no stars respectively from literature (Masseron et 
al. 2010). Black open squares, open circles represent CH giants 
and subgiants respectively from literature (Karinkuzhi \& Goswami 2014, 
2015, Goswami et al. 2006, 2016, Sneden \& Bond 1976, Vanture 1992, 
Goswami \& Aoki 2010, Jonsell et al. 2006, Masseron et al. 2010). 
HD 36650 (solid triangle), HD 207585 (six sided cross), HD 219116 
(solid square).}}\label{abundance comparison}
\end{figure}

\noindent Detailed chemical abundance analysis of the program stars 
HD 36650, HD 207585  and HD 219116 were carried out. The absolute abundances 
and the abundance ratios of light elements- Na and Mg as well as neutron-capture
elements such as Sr, Y, Zr, Ba, La, Ce, Nd, and- Sm are 
determined for these stars. The observed abundance ratios are compared 
with their counterparts 
in other barium stars from literature. The comparison is also done with the CH stars,
CEMP stars and normal field giants from  literature. 
Figure \ref{abundance comparison} illustrates such a comparison for a few neutron-capture 
elements. As it is evident from the figure, the neutron capture elements show enhanced 
abundances compared to normal field giants of similar metallicity.

From the location of the stars in the HR diagram (log T$_{eff}$ v/s log (L/L$_{\odot}$) plot), 
it is clear that HD 207585 is on the subgiant branch (SGB) and the stars HD 36650 and HD 219116 
are on the first ascent of giant branch (RGB). 
Though the star HD 207585 is classified in SIMBAD  as a 
luminosity class II star,
our analysis shows that this object is a subgiant star with log g value 3.80.
The estimate of log g using the parallax method gives a value 3.90 which
is consistent with the spectroscopic estimate. Our estimate is close to the 
literature values 4.00 (Masseron et al. 2010) and 3.50 (Smith \& Lambert 1986a).  
At these evolutionary stages 
of these stars,  the s-process nucleosynthesis is not likely to happen and 
hence it is likely that the observed overabundance of heavy
elements are extrinsic. A binary mass-transfer scenario, in which the 
products  of s-process nucleosynthesis from a companion (now invisible 
white dwarf) AGB star contaminated  the surface of the now observed 
barium star could explain this observed overabundance. 
According to Jorissen et al. (1998), binarity is a necessary 
condition to produce Ba stars, but it is not a sufficient condition.

We have also estimated the abundance ratios [ls/Fe], [hs/Fe] and [hs/ls], 
where ls represents  light s-process elements Sr, Y and Zr, and hs 
represents heavy s-process elements Ba, La, Ce and Nd. 
The [hs/ls] values are given in  Table \ref{important abundance ratios}.
[hs/ls] is an indicator of s-process efficiency. As the metallicity decreases, 
the number of seed nuclei for neutron-capture decreases, and more 
neutrons per seed nuclei will be available. This favours 
the production of heavy s-process elements. As a result, the [hs/ls] 
ratio increases with decreasing metallicity (Clayton 1988, Wallerstein 1997, 
Goriely \& Mowlavi 2000, Busso et al. 2001, de Castro et al. 2016, Cristallo et al. 2011). 
According to the AGB models of Busso et al. (2001), at metallicities higher than solar,
the light s-process elements (ls) are the dominant product over heavy s-process elements (hs). 
At lower metallicities, hs are the dominant products over ls.  
All these three stars show [hs/ls] $>$ 0, indicating the overabundance
of Ba-peak elements over Sr-peak elements. Our estimates 
of [hs/ls] agree with the model calculations of Busso et al. (2001) 
for similar metallicities.

\begin{table}
\begin{center}
\caption{\textbf{[hs/ls] ratios of program stars}} \label{important abundance ratios}
\begin{tabular}{|l |l |l |l |l |}
\hline\hline 
            &           &          &          &           \\
Star name   & [Fe/H]    & [ls/Fe]  & [hs/Fe]  & [hs/ls]   \\ 
            &           &          &          &           \\
\hline
HD 36650    & $-$0.02     & 0.62     & 0.84     & 0.22       \\
HD 207585   & $-$0.38     & 1.29     & 1.66     & 0.37       \\
HD 219116   & $-$0.45     & 0.71     & 1.32     & 0.61       \\
\hline
\end{tabular}
\end{center}
 \end{table}

The $^{13}$C($\alpha$, n)$^{16}$O reaction is the major 
neutron source in low-mass AGB stars with initial mass $\leq 3M_{\odot}$. 
The neutron density produced by this source is $\sim$10$^{8}$ 
neutrons/cm$^{3}$ which is much lower than the density, $\sim$10$^{13}$ 
neutrons/cm$^{3}$, produced by $^{22}$Ne($\alpha$, n)$^{25}$Mg (Busso et 
al. 2001, Goriely \& Mowlavi 2000). The efficiency of the neutron source 
$^{13}$C($\alpha$, n)$^{16}$O is anti-correlated with metallicity 
(Clayton 1988, wallerstein 1997). Hence low [hs/ls] ratios can be 
expected for the low mass stars at near solar metallicities considering 
the neutron source $^{13}$C($\alpha$, n)$^{16}$O. Also, massive AGB stars 
with masses between 5 and 8 $M_{\odot}$ can show low [hs/ls] ratios 
considering the neutron source $^{22}$Ne($\alpha$, n)$^{25}$Mg 
(Busso et al. 2001, Karakas \& Lattanzio 2014).  But in massive AGB stars, 
Na and Mg are strongly produced as a result of $^{22}$Ne burning 
(Bisterzo et al. 2010). But none of our stars show enhanced abundances 
of Na and Mg. This may indicate the operation of the $^{13}$C source 
in the former AGB companion, which in turn suggests a low mass for 
the companion. 

Another indicator of neutron density and mass of AGB stars is the 
ratio [Rb/Sr]. Neutron density required for the production of Rb 
from Kr is $\sim$ 5$\times$10$^{8}$ neutrons/cm$^{3}$ indicating 
the operation of the $^{22}$Ne($\alpha$, n)$^{25}$Mg reaction. 
In low-mass stars, Sr is produced instead. Thus the ratio [Rb/Sr]
should be positive in massive AGB stars (Karakas et al. 2012). 
We have searched for Rb lines in our spectra. The Rb I line 
at 7947.597 {\rm \AA} returns a [Rb/Sr] value $-$0.56 in HD 36650. 
We could not determine the Rb abundance in other stars as no 
useful lines were detected in their spectra for the abundance analysis. 

\begin{figure}
\begin{center}
\includegraphics[width=.54\textwidth,height=.54\textwidth]{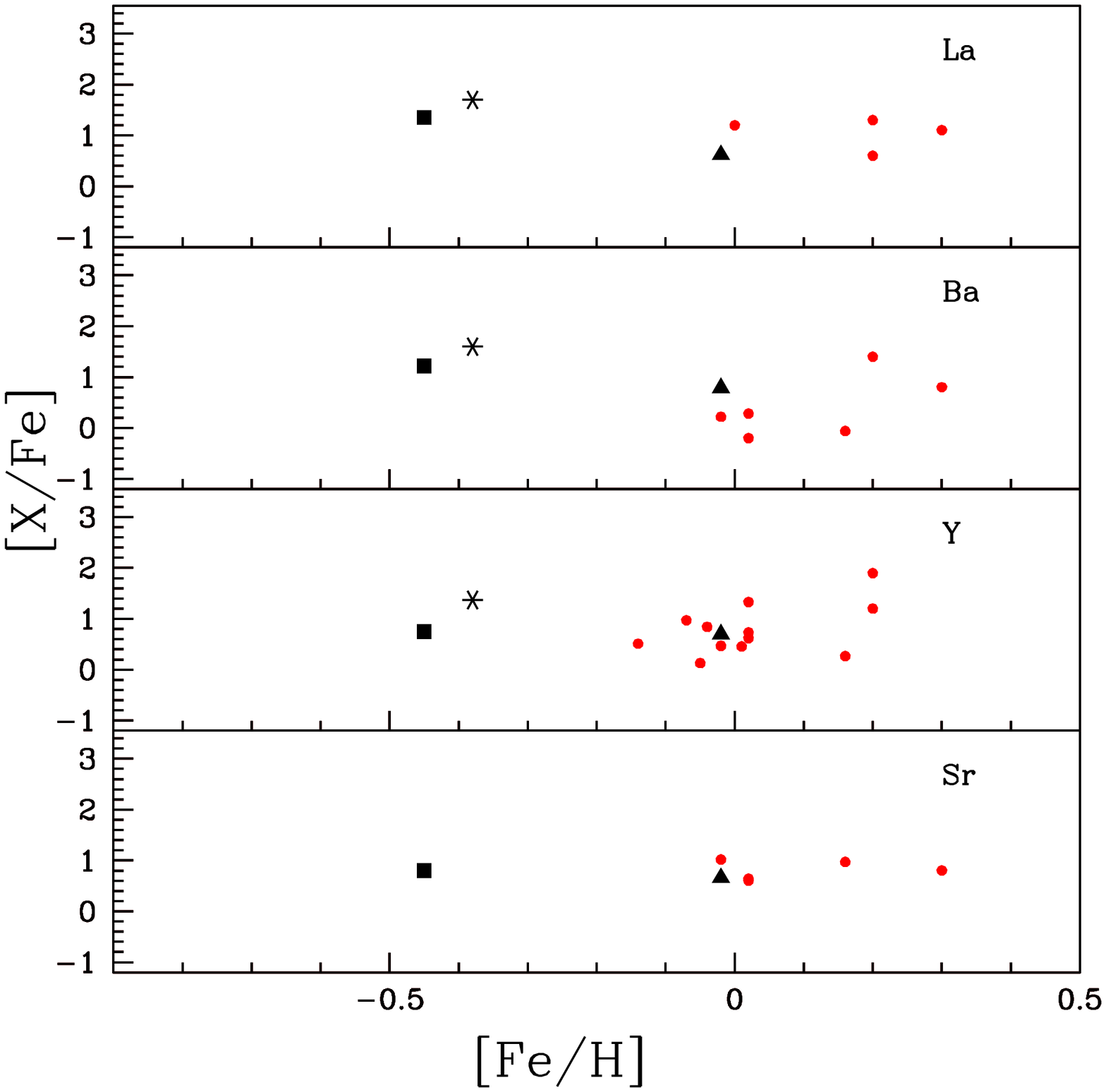}
\caption{\small{Comparison of abundance ratios of neutron-capture elements 
observed in the program stars and the AGB stars with respect to metallicity 
[Fe/H]. Red circles represent the AGB stars from literature 
(Smith \& Lambert 1985, 1986b, 1990, Abia \& wallerstein 1998).}}\label{AGB comparison}
\end{center}
\end{figure}

We have compared the neutron-capture elemental abundance ratios of the 
program stars with that of low-mass AGB stars from literature 
where the $^{13}$C source is operating. The comparison is shown in  
Figure \ref{AGB comparison}. The abundance ratios are found to match
closely. The scatter in the ratio may be a consequence of different 
dilution factors during the mass transfer as well as the orbital 
parameters, metallicity and initial mass (de Castro et al. 2016). 

\section{Conclusions}
We have conducted a detailed 
spectroscopic analysis for a  sample of three barium 
stars. The abundance estimates are consistent with the operation 
of the $^{13}$C($\alpha$, n)$^{16}$O source in the former low-mass AGB companion.
Detection of Rb I line at 7947.597 {\rm \AA} in the  spectrum of HD 36650 
allowed us to determine the [Rb/Sr] ratio with a value of $-$0.56. This ratio 
indicates the operation of 
$^{13}$C($\alpha$, n)$^{16}$O in the companion star. As this reaction 
occurs in the low-mass AGB stars, we confirm that the former companion 
of HD 36650 is a low-mass AGB star. 
Distribution of abundance patterns and [hs/ls] ratios indicate 
low-mass companions for the objects HD 207585 and HD 219116. 
While in this work we have presented the results from  the analyses of three
objects, a detailed study including a larger sample is in preparation.

\section*{Acknowledgements}
The authors would like to acknowledge the organizers of the 2$^{nd}$ BINA 
workshop for the local hospitality and the financial support. SJ would 
also like to acknowledge IIA for partial financial support. Funding 
from DST SERB project No. EMR/2016/005283 is gratefully acknowledged. AG 
would also like to acknowledge the financial support  from BINA projects 
DST/INT/Belg/P-02 (India) and BL/11/IN07 (Belgium).

\footnotesize
\beginrefer

\refer Abia C., Wallerstein G. 1998, MNRAS, 293, 89

 \refer Allen D. M.,  Barbuy B. 2006, A\&A, 454, 895 
 
 \refer Alonso A., Arribas S., Martinez-Roger C. 1999, A\&ASS, 140, 261 
 
 \refer Asplund M., Grevesse N., Sauval A.J. et al. 2009, ARA\&A, 47, 481
  
 \refer Barbuy B., Jorissen A., Rossi S. C. F. et al. 1992, A\&A, 262, 216 
 
 \refer Bartkevicius A. 1996, Balt. Astron., 5, 217 
 
 \refer Bisterzo S., Traglio C., Wiescher M. et al. 2010, MNRAS, 404, 1529 
 
 \refer Bidelman W.P.,  Keenan P.C. 1951, ApJ, 114, 473 
 
 \refer Busso M., Gallino R., Wasserburg G. J. 1999, ARA\&A, 37, 239 
 
 \refer Busso M., Gallino R., Lambert D. L. et al. 2001, ApJ, 557, 802 
 
 \refer Cayrel R., Depagne E., Spite M. et al. 2004, A\&A, 416, 1117
 
 \refer Chen B., Vergely J.L, Valette B. et al. 1998, A\&A, 336,137 
 
 \refer Clayton D. D. 1988, MNRAS, 234, 1 
 
 \refer Cristallo S., Piersanti L., Sraniero O. et al. 2011, ApJS, 197,17 
 
 \refer de Castro D.B., Pereira C.B., Roig F. et al. 2016, MNRAS, 459, 4299 
 
 \refer Drake N. A., Pereira C. B. 2008, AJ, 135, 1070 

 \refer Dray L. M., Tout C. A., Karakas A. I. et al. 2003, MNRAS, 338, 973
 
 \refer Girardi L., Bressan A., Bertelli G. et al. 2000, A\&AS, 141, 371  
 
 \refer Goriely S., Mowlavi N. 2000, A\&A, 362, 599 
 
 \refer  Goswami A., Aoki W., Beers T. C. et al. 2006 ,MNRAS, 372, 343 
  
 \refer Goswami A., Aoki W. 2010, MNRAS, 404, 253 
 
 \refer Goswami A., Aoki W., Karinkuzhi D. 2016, MNRAS, 455, 402 

\refer Henry R. B. C., Edmunds M. G., K\"oppen J. 2000, ApJ, 541, 660 
 
 \refer Jonsell K., Barklem P. S., Gustafsson B. et al. 2006, A\&A, 451, 651 
 
 \refer Jorissen A., Van Eck S., Mayor M. et al. 1998, A\&A, 332, 877 
 
 \refer Jorissen A., Boffin H. M. J., Karinkuzhi D. et al. 2019, arXiv:1904.03975
 
 \refer Karakas A. I., Garcia-Hernandez D. A., Lugaro M. 2012, ApJ, 751, 8 
 
 \refer Karakas A. I., Lattanzio J. C. 2014, PASA, 31, 30 
 
 \refer Karinkuzhi D., Goswami A. 2014, MNRAS, 440, 1095 
 
 \refer  Karinkuzhi D., Goswami A. 2015, MNRAS, 446, 2348 
 
 \refer Lu P. K. 1991, AJ, 101, 2229 
 
 \refer Luck R. E., Bond H. E. 1991, ApJS, 77, 515 
 
 \refer Luck R. E., Heiter U. 2007,AJ, 133, 2464 
 
 \refer Masseron T., Johnson J. A., Plez B. et al. 2010, A\&A, 509, A93 
 
 \refer McClure R. D. 1983, ApJ, 268, 264 
 
 \refer McClure R. D. 1984, PASP, 96, 117 
 
 \refer McClure R. D., Woodsworth W. 1990, ApJ, 352, 709 
 
 \refer McClure R. D., Fletcher J. M., Nemec J. 1980, ApJ, 238, L35 
 
 \refer McWilliam A. 1998, AJ, 115, 1640 
 
 \refer Pereira C. B., Drake N. A. 2009, A\&A, 496, 791
 
 \refer Prochaska J. X., McWilliam A. 2000, ApJ, 537, L57 
 
 \refer Prochaska J. X., Naumov S. O., Carney B. W. et al. 2000, AJ, 120, 2513
 
 \refer Smith V. V., Lambert D. L. 1985, ApJ, 294, 326
 
 \refer Smith V. V., Lambert D. L. 1986a, ApJ, 303, 226
 
 \refer Smith V. V., Lambert D. L. 1986b, ApJ, 311, 843
 
 \refer Smith V. V., Lambert D. L. 1990, ApJS, 72, 387
 
 \refer Sneden, C. 1973, Ph.D. Thesis, Univ. of Texas 
 
 \refer Sneden C., Bond H. E. 1976, ApJ, 204,810 
 
 \refer Udry S., Jorissen A., Mayor M. et al. 1998a, A\&AS, 131, 25 
 
 \refer Udry S., Mayor M., Van Eck S. et al. 1998b, A\&AS, 131, 43 
 
 \refer Vanture A. 1992, AJ, 104, 5  
 
 \refer Wallerstein G. 1997, RvMP, 69,995 
 
 \refer Worely C.C., Hill V. J., Sobeck J. et al. 2013, A\&A, 553, A47 
 
 \refer Yang G. C., Liang Y. C., Spite M. et al. 2016, RAA, 16, 19 
 
\endrefer           

\end{document}